\documentclass[twocolumn,showpacs,preprintnumbers,amssymb]{revtex4}

\usepackage{graphicx}
\usepackage{epsfig}
\usepackage{dcolumn}
\usepackage{bm}
\def\gsim{\lower0.5ex\hbox{$\:\buildrel >\over\sim\:$}}
\def\lsim{\lower0.5ex\hbox{$\:\buildrel <\over\sim\:$}}

\let\a=\alpha\let\b=\beta

\newcommand{\be}{\begin{equation}}
\newcommand{\ee}{\end{equation}}
\newcommand{\bea}{\begin{eqnarray}}
\newcommand{\eea}{\end{eqnarray}}

\newcommand{\nbox}{{\,\lower0.9pt\vbox{\hrule \hbox{\vrule height 0.2 cm
\hskip 0.2 cm \vrule height 0.2 cm}\hrule}\,}}

\begin{document}

\preprint{UCI-TR-2007-47}

\title{Models and Phenomenology of Maximal Flavor Violation}

\author{Shaouly Bar-Shalom$^{a,b}$}
\email{shaouly@physics.technion.ac.il}
\author{Arvind Rajaraman$^b$}
\email{arajaram@uci.edu}
\affiliation{$^a$ Physics Department, Technion-Institute of Technology, Haifa 32000, Israel\\
$^b$Department of Physics and Astronomy, University of California, Irvine, CA 92697, USA}

\date{\today}

\begin{abstract}
We consider models of maximal flavor violation (MxFV),
in which a new scalar mediates large $q_3 \leftrightarrow q_1$ or $q_3 \leftrightarrow q_2$
flavor changing transitions ($q_i$ is an $i$'th generation quark),
while $q_3 \leftrightarrow q_3$ transitions are suppressed, e.g.,
$\xi_{31}, ~\xi_{13} \sim V_{tb}$ and $\xi_{33} \sim V_{td}$,
where $\xi_{ij}$ are the new scalar couplings to quarks and $V$ is the CKM matrix.
We show that, contrary to the conventional viewpoint,
such models are not ruled out by the existing low energy data on
$K^0,~B^0$ and $D^0$ oscillations and rare $K$ and $B$-decays.
We also show that these models of MxFV can have surprising new signatures
at the LHC and the Tevatron.
\end{abstract}


\maketitle

{\it Introduction:}
The hierarchy problem and the problem of dark matter imply that
the Standard Model (SM) is incomplete. Models of new physics (NP) typically
predict a large number of new particles at the TeV scale, and in
particular, many models predict the existence of new scalars
beyond the SM Higgs, which, in principle, can induce flavor violating (FV)
currents through Yukawa-like interactions of the form
$\xi_{ij} \Phi \bar{q}_i q_j$ \cite{ARS,sonilunghi}.
It is traditionally assumed that, to satisfy the precision data,
the new particles that mediate FV have to be very
heavy (i.e., $\sim 10$ TeV - somewhat in conflict with
the solutions to the hierarchy problem and dark matter) or
else their FV couplings have to be very small.
In general, this leads to the ``NP flavor puzzle'' since
if there is NP at the TeV scale, then there is no underlying principle that would forbid
low energy FV signatures at large rates.

A popular solution to
the flavor problem is to
impose the
minimal FV (MFV) ansatz, which states that the NP
is ``aligned'' with the SM, such that 
all FV transitions are governed by the nearly diagonal CKM matrix $V$.
For instance, since the CKM elements satisfy e.g. $V_{td},~V_{ts} \ll V_{tb}$, the MFV
ansatz imposes the couplings of any new scalar to a pair of top+light quark to satisfy
$\xi_{3i}, \xi_{i3} \ll \xi_{33}$ for $i=1,2$.
As an example,
the Minimal Supersymmetric SM (MSSM) soft breaking
terms allow in general for ${\cal O}(1)$
FV sfermion mixings which can mediate unacceptably large
contributions to FV processes in meson mixings and decays \cite{0710.5206}.
In this case sfermion-fermion alignment which leads to MFV is
motivated since otherwise the sfermion masses have to be
too heavy for addressing the fine-tuning problem.

In this note, we
present a new class of scalar mediated FV models,
which maximally depart from the MFV ansatz,
(in the sense that $\xi_{31},\xi_{32} \sim {\cal O}(1) \gg \xi_{33}$)
and still satisfy all constraints from flavor physics even with a relatively light scalar
 $m_{\Phi} \sim m_W$.
To emphasize the contrast with MFV models,
we will describe these models as having Maximal FV (MxFV).
We can  further subdivide these MxFV models  into MxFV$_1$ models
where $\xi_{31},\xi_{13} \sim {\cal O}(1)$ while $\xi_{32},\xi_{23} \ll
{\cal O}(1)$  and MxFV$_2$ models if $\xi_{32},\xi_{23} \sim
{\cal O}(1)$ while $\xi_{31},\xi_{13} \ll {\cal O}(1)$, and in both cases $\xi_{33},~\xi_{ij} \sim 0$ $i,j=1,2$.
We show that such models may be as viable as the MFV models.

We will begin by presenting a model independent setup where
MxFV is mediated by a new scalar with a mass of a few hundred GeV.
We then systematically analyze constraints on these
models coming from precision data on meson mixing and decays and
show that, contrary to expectations, we can indeed have electroweak scale
extensions of the SM that fall into the MxFV categories.
Finally, we show that such models of MxFV can have
surprising new signatures at hadron
colliders (LHC \& Tevatron). We close with a discussion of future
directions.

{\it Models of Maximal Flavor Violation:}
In what follows, we will consider the more restrictive case (with respect to having
tighter constraints from flavor physics) where
the new scalar is
a doublet of electroweak $SU(2)$.
However, the analysis below can be applied to singlets as well.

Let $\Phi_{FV}$ be a new scalar doublet that potentially mediates MxFV through:
\bea
{\cal L}_{FV} = \xi^u_{ij}  \bar Q_{iL} \tilde\Phi_{FV} 
u_{jR} +
\xi^d_{ij}  \bar Q_{iL} \Phi_{FV}  d_{jR} + h.c.~, \label{LFV}
\eea
\noindent
where $i,j$ are generation indices.
Now, flavor changing neutral interactions in the down-quark sector are severely constrained by
meson-mixings \cite{ARS}, implying in general that $\xi^d \ll \xi^u$.
We will therefore set $\xi^d\sim 0$ and consider
only effects from $\xi^u \equiv \xi$.
Defining $\Phi_{FV} = (H^+ , H^0)$
(note that in general $H^0$ is complex) we can then re-write (\ref{LFV}) as:
\begin{eqnarray}
{\cal L}_{FV} \sim \xi_{ij} \cdot \left( H^{0} \bar u_{iR}  u_{jL} -
H^+ \bar u_{iR}  d_{jL} \right)  + h.c.~, \label{xicoup}
\end{eqnarray}
where e.g., $\xi_{31}$ is the coupling of $H^+ td$ and by SU(2) also of $H^0 t u$.

We wish to construct models where e.g., $\xi_{31},~\xi_{32} \gg \xi_{33}$.
We can do that by imposing flavor symmetries. One simple example is
a $Z_2$ symmetry under which the SM Higgs and the 1st and 2nd generation quarks are even
while $\Phi_{FV}$ and the 3rd generation quarks are odd:
\begin{eqnarray}
 &&\Phi_{SM} \to \Phi_{SM}~,~q_{i} \to q_{i} ~,i=1,2 ~, \nonumber \\
 &&\Phi_{FV} \to - \Phi_{FV}~,~q_3 \to -q_3 \label{Z2}~,
\end{eqnarray}
This symmetry suppresses the CKM elements $V_{td},V_{ts},V_{ub},V_{cb}$ and
simultaneously suppresses the new $H^+ t b$ and $H^0 t t$
interactions ($\xi_{33}$), therefore
under this $Z_2$ symmetry we have
\begin{eqnarray}
\xi \equiv \begin{pmatrix} {
0 & 0 & \xi_{13} \cr
0 & 0 & \xi_{23} \cr
\xi_{31} & \xi_{32} & 0 } \end{pmatrix}
\label{texture}~.
\end{eqnarray}
If the $Z_2$ is broken weakly (e.g., by a very small $\Phi_{FV}$ condensate or by higher dimensional
operators), then
a small value for the CKM elements $V_{td},V_{ts},V_{ub},V_{cb}$
as well as for all zero entries in (\ref{texture}) are
generated. We then expect 
$\xi_{33} \sim V_{td}~,V_{ts}$,
while maintaining $\xi_{31},~\xi_{32} \sim V_{tb} \sim {\cal O}(1)$.

Another possibility is to have U(1) flavor symmetries. For example, if the three fermion generations
have charges $\a,\a+1,\b$ respectively, and $\Phi_{FV}$ has a charge $\a-\b$, then
the only tree level coupling of the new Higgs is $\xi_{31}$ while all the other
couplings are suppressed.
The SM Higgs has charge 0 under the U(1), thus allowing all diagonal mass terms.
When the U(1) is weakly broken, we can expect all couplings to be generated, but we will still have
$\xi_{31} \gg \xi_{33}$ and $V_{tb} \sim \xi_{31}$, $V_{td} \sim \xi_{33}$.

Here, we will not restrict ourselves to any model, and
instead take a phenomenological approach towards the general MxFV case,
where we assume non-zero $\xi_{31},\xi_{32},\xi_{13},\xi_{23}$ and $\xi_{33} \sim 0$ as well
as $\xi_{ij} \sim 0$ for $i,j=1,2$.
We then consider the experimental constraints on such textures
and show that while some products of the above couplings are bounded to be rather small,
the MxFV$_{1,2}$ scenarios are not ruled out.

The phenomenology of these models depends sensitively on the details of the scalar sector.
In general one can have mixing terms between
the SM Higgs and the new doublet e.g. a term like $\lambda D_\mu \Phi_{SM} D^\mu \Phi_{FV}$.
After the SM Higgs develops a VEV this term can lead to interactions like
$\sim \lambda m_W W^+ W^- H^0$, which make the phenomenology difficult to analyze without
knowing the size of $\lambda$.
We shall assume that such mixings terms are absent since we expect
them to be very small; indeed, in the
models discussed above, the flavor symmetries prohibit such terms, and when they are weakly
broken we expect $\lambda \sim \xi_{33} \sim V_{td} \ll 1$.
Note that a $WWH^0$ interaction
term can also arise from the kinetic term $|D_\mu \Phi_{FV}|^2$,
if the flavor symmetry is broken by a small VEV of $\Phi_{FV}$.
We shall also ignore the couplings of the new Higgs doublet to leptons.

{\it Experimental Constraints:}
We now consider the existing experimental constraints on the MxFV
scenarios described above. For simplicity
we will assume that all the new FV couplings are real.
We find that the most stringent constraints come from $F^0 - \bar F^0$ mixings, $F=K,~B_d,~B_s,~D$ and
from rare $K$-decays. Limits from $B$ decays (semi-leptonic or
hadronic) are either weaker or depend on the couplings of $\Phi_{FV}$ to leptons.
As will be shown below, the above observables constrain only the following combinations of
couplings:
\begin{eqnarray}
 \eta_K \equiv \xi_{31} \cdot \xi_{32} ~~&,&~~ \eta_D \equiv \xi_{13} \cdot \xi_{23} ~, \nonumber \\
 \eta_{B_d}^1 \equiv \xi_{31} \cdot \xi_{23} ~~&,& ~~\eta_{B_d}^2 \equiv \xi_{31} \cdot \xi_{13} ~, \nonumber \\
 \eta_{B_s}^1 \equiv \xi_{32} \cdot \xi_{23} ~~&,& ~~\eta_{B_s}^2 \equiv \xi_{32} \cdot \xi_{13}
 \label{FVcouplings}~.
\end{eqnarray}
\noindent

\noindent\underline{$K^0 - \bar K^0$:}

The $K^0 - \bar K^0$ mass difference and
measure of indirect CP-violation in the K system
are given by (see e.g. \cite{9806471}):
\begin{eqnarray}
 \Delta m_K = 2 {\rm Re}(M_{12}^K)~,~\epsilon_K \equiv \frac{exp(i\pi/4)}{\sqrt{2} \Delta m_K}
 {\rm Im}(M_{12}^K)~.
\end{eqnarray}

The MxFV contribution to the above observables
arises from new $WH$ and $HH$ box diagrams ($H$ stands for the charged scalar), where only the
top-quark can propagate in the loops. These new box diagrams shift
$M_{12}^K$ by (the contribution of a generic new scalar
to meson mixings can be found in \cite{0107048}):
\begin{eqnarray}
 \delta M_{12}^{MxFV} &=& \frac{m_K f_K^2 \hat B_K}{12 \pi^2 m_W^2} \frac{\eta_2}{4r}
 \left( \frac{2 \pi \alpha}{\sin^2\theta_W} V_{td}^\star V_{ts} \eta_K f(x_t,x_H) \right.
 \nonumber \\
 &+& \left. \frac{\eta_K^2}{8} g(x_t,x_H) \right) ~,
\end{eqnarray}
\noindent where (see e.g. \cite{sonilunghi})
we use $m_K=0.498$ GeV, $f_K=0.159$ GeV, $\hat B_k = 0.79$, $\eta_2=0.57$, $r=0.985$ and
\begin{eqnarray}
 f(x_t,x_H) &=& \frac{x_t}{x_t-x_H} \left(\frac{\log(x_t)}{x_t-1} + \frac{x_H \log(x_H)}{(x_H-1)(x_t-x_H)} \right)
 \nonumber~,\\
 g(x_t,x_H) &=& \frac{1}{(x_t-x_H)^2} \left( x_H + 2 x_t \log\left(\frac{x_t}{x_H}\right) \right)  ~,
\end{eqnarray}
\noindent where $x_i \equiv m_i^2/m_W^2$.

\begin{figure}[htb]
\epsfig{file=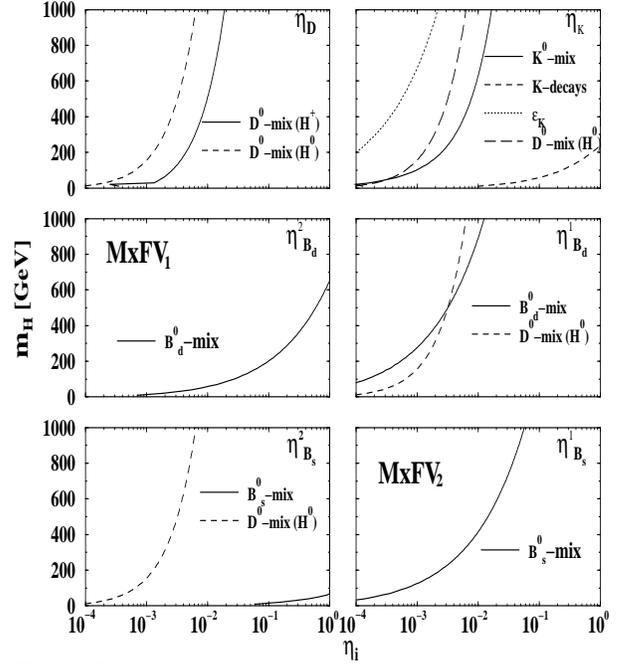,height=9cm,width=8cm,angle=0}
\vspace{-0.6cm}
\caption{\emph{Portions of the $\eta_i - m_{H^+}$ planes and of the $\eta_i - m_{H^0}$ planes
(when indicated by $D^0$-mix ($H^0$)) excluded by the various K and B-physics low energy observables.
In each sub-figure the relevant FV coupling product $\eta_i$ is indicated on the upper right corner and the
area below the curves is excluded.}}
\label{fig1}
\end{figure}

In Fig.~\ref{fig1} we plot the allowed range in the $\eta_K - m_{H^+}$ plane
demanding that
(i) the contribution of ${\rm Re}(\delta M_{12}^{MxFV})$ to $\Delta m_K$ does not exceed the long distance
contributions estimated to be less than $\sim 0.3 \Delta m_K^{exp}$, where
$\Delta m^{exp}_K = 0.005301$ ps$^{-1}$ is the observed mass splitting \cite{PDG},
and (ii) the contribution of ${\rm Im}(\delta M_{12}^{MxFV})$
to $\epsilon_K$ does not exceed
$ 0.4 \epsilon_K^{exp}$, where $\epsilon_K^{exp} = 2.232 \cdot 10^{-3}$ is the central measured value
\cite{PDG}. In particular, we have traded uncertainties in the theoretical input for $\epsilon_K$ for
a generous $40\%$ deviation from the central observed value. Note that
${\rm Im}(\delta M_{12}^{MxFV}) \propto {\rm Im}(V_{td}^\star V_{ts})$ since we are taking
all $\eta_i$ in (\ref{FVcouplings}) to be real.

\noindent\underline{Rare $K$-decays:}

The rare K-decays $K^+ \to \pi^+ \nu_L^\ell \bar\nu_L^\ell$ and
$K^0_L \to \pi^0 \nu_L^\ell \bar\nu_L^\ell$ (and other
$K \to \pi$ decays that go through the $(\bar s d)_{V-A}$ current) are mediated by the following effective Hamiltonian \cite{9806471,rareK}:
\begin{eqnarray}
 {\cal H}_{eff}=\frac{G_F}{\sqrt{2}} \frac{2 \alpha}{\pi s_W^2} \left(X_{SM} + \delta X^{NP} \right)
 \cdot \left(\bar s_L \gamma_\mu d_L\right)\left(\bar\nu_L^\ell \gamma_\mu \nu_L^\ell \right)
\end{eqnarray}
\noindent where $X_{SM} \sim 8.3 \cdot 10^{-4}$ is the SM part and $\delta X^{NP}$
is the new physics contribution.
In the MxFV framework, there are new Z-penguin 1-loop diagrams with $H^+-t$ exchanged in the loops,
which give:
\begin{eqnarray}
 \delta X^{MxFV} = \eta_K \cdot \frac{s_W^2}{32 \pi \alpha} \frac{y_t}{1-y_t}
 \left(1 + \frac{\log(y_t)}{1-y_t} \right)~,
\end{eqnarray}
\noindent where $y_t \equiv m_t^2/m_{H^+}^2$.

In Fig~\ref{fig1} we plot the
allowed range in the $\eta_K - m_{H^+}$ plane by imposing $\delta X^{MxFV} < X_{SM}$.

\noindent\underline{$B_q^0 - \bar B_q^0$:}

Let us define (q=d or s)
\begin{eqnarray}
 R_q \equiv \frac{\Delta m_q}{\Delta m_q^{SM}} \equiv 1 + \delta_q^{NP} \label{Rq}~,
\end{eqnarray}
where $\Delta m_q \equiv \Delta m_q^{SM} + \delta m_q^{NP}$ is the total $B_q^0 - \bar B_q^0$ mass splitting
and $\delta_q^{NP} \equiv \delta m_q^{NP}/\Delta m_q^{SM}$ represents the portion of NP.

Taking the experimentally observed
values $\Delta m_s^{exp} = 17.77 \pm 0.17$ ps$^{-1}$ \cite{delmsexp}
and $\Delta m_d^{exp} = 0.507 \pm 0.004$ ps$^{-1}$ \cite{delmdexp}, and
the SM predictions (averaging between the two HP and JL groups \cite{delmsth1}
and adding the errors in quadrature)
$\Delta m_s^{SM} = 18.7 \pm 4.3$ ps$^{-1}$ and $\Delta m_d^{SM} = 0.69 \pm 0.14$ ps$^{-1}$, we obtain
$R_d =0.74 \pm 0.15$ and $R_s =0.95 \pm 0.22$.
We then require that the MxFV contribution
is within 1 standard deviation of the
central values of $|R_q -1|$:
\begin{eqnarray}
|\delta_s^{MxFV}| <  0.27 ~~ \& ~~ |\delta_d^{MxFV}| <  0.41 \label{limB}~.
\end{eqnarray}

In the MxFV framework the $B_q^0 - \bar B_q^0$ mass splitting receives a 
contribution from
new $WH$ box diagrams with an exchange of either $t$ and $c$ or $t$ and $u$ quarks in the loop
(there are no $HH$ box diagrams within the MxFV setup), which give:
\begin{eqnarray}
 \delta_q^{MxFV} = - \frac{s_W^2 r}{4 \pi \alpha} \cdot \frac{V_{cq} \eta_{B_q}^1 \tilde f(x_c) +
V_{uq} \eta_{B_q}^2 \tilde f(x_u) }{V_{tb} (V_{tq}^\star)^2 S_0(x_t) }  ~,
\end{eqnarray}
\noindent where $S_0(x_t) \simeq 2.46 \cdot [m_t/170 ~ {\rm GeV}]^{1.52}$ and
\begin{eqnarray}
 \tilde f(x_i) &\simeq& \sqrt{x_t x_i} \left( \frac{\log(x_H)}{(x_H-1)(x_H-x_t)}
 \right. \nonumber \\
 &-&\frac{\log(x_t)}{(x_t-1)(x_H-x_t)}
 - \left. \frac{x_i}{x_H x_t} \log(x_i) \right)~.
\end{eqnarray}

In Fig.~\ref{fig1} we plot the
allowed range in the $\eta_{B_q}^i - m_{H^+}$ planes by imposing (\ref{limB}).

\noindent\underline{$D^0 - \bar D^0$:}

In the case of $D^0$ oscillations there are two types of 1-loop box diagrams that contribute
due to the new MxFV interaction Lagrangian in (\ref{xicoup}): the charged Higgs
$WH$ and $HH$ boxes with b-quark exchanges and the neutral Higgs $H^0 H^0$ boxes with top-quark
exchanges, where $H^0$ corresponds in general to any one of the two (CP-odd or CP-even) neutral
Higgs components of $\Phi_{FV}$.
We find that the following expression is
a good approximation
for the $D^0 - \bar D^0$ mass difference generated by these MxFV box diagrams:
\begin{eqnarray}
 \delta M_D^{MxFV} \sim \frac{f_D^2 m_D}{16 \pi^2} \frac{A_D}{m_W^2} \cdot
 \left (\delta^{WH}+\delta^{HH}+\delta^{H_i^0}_1 + \delta^{H_i^0}_2 \right)
\end{eqnarray}
\noindent where we have absorbed all non-perturbative $\hat B_i$ parameters and additional
numeric factors into $A_D$ which is then of ${\cal O}(1)$ \cite{DD-mix}. For the various
$\delta$'s we get:
\begin{eqnarray}
\delta^{WH} &\sim& \eta_D V_{ub}^\star V_{cb} \frac{2 \pi \alpha}{s_W^2} \frac{x_b}{x_{H^+}}
\left( \log(x_b) - \frac{\log(x_{H^+})}{1-x_{H^+}} \right) \\
\delta^{HH} &\sim& \frac{\eta_D^2}{8} \frac{1}{x_{H^+}}
\left( 1 + 2 \frac{x_b}{x_{H^+}} \log\left(\frac{x_b}{x_{H^+}}\right) \right) ~,
\end{eqnarray}
and assuming that only one of the neutral components dominates the neutral Higgs exchanges (i.e.,
$H^0 = H_1^0$ and $m_{H_2^0} \gg m_{H_1^0}$) we have:
\begin{eqnarray}
 \delta^{H^0}_1 &\sim&  
 \frac{\left( \eta_{B_d}^1 + \eta_{B_s}^2 \right)^2
 x_t}{(x_t-x_{H^0})^2} \left( 1 + \log \left(\frac{x_t}{x_{H^0}}\right) \right) ~, \\
 \delta^{H^0}_2 &\sim& 
 \frac{\left( \eta_K + \eta_D \right)^2
 x_t}{(x_t-x_{H^0})^2} \left( \frac{x_{H^0}}{x_t} + 2 \log \left(\frac{x_t}{x_{H^0}}\right) \right) ~,
\end{eqnarray}

In the case of the $D^0 - \bar D^0$ mass splitting the SM prediction is
dominated by long distance effects. Therefore, in deriving the limits on the FV coupling products
that enter $D^0$ oscillations, we require $\delta M_D^{MxFV}$ not to exceed the
measured value $\Delta m_D^{exp} = (14.5 \pm 5.6) \cdot 10^{-3}$ ps$^{-1}$ \cite{DDmixexp}.
In Fig.~\ref{fig1} we plot the
allowed range in the $\eta_D - m_{H^+}$ plane and $\eta_i - m_{H^0}$ planes (i.e.,
$\eta_i= \eta_{B_d}^1,~\eta_{B_s}^2,~\eta_K,~\eta_D$ from $H^0$ exchanges) imposed by
$\delta M_D^{MxFV} < 20.1$ ps$^{-1}$.

{\it Results and Discussion:} Our analysis leads to the following results;

1. Any MxFV model in which only one FV
coupling is of ${\cal O}(1)$ while all others are $\sim 0$
is not constrained
by meson flavor physics, regardless of $m_{\Phi_{FV}}$.

2. In MxFV$_1$ models where the new FV Higgs field couples only to the third and first generations,
the only constraint comes from $B_d^0 - \bar B_d^0$ mixings. This constraint is, however,
rather weak as it allows e.g., $m_{H^+},~m_{H^0} \sim 200$ GeV
even with ${\cal O}(1)$ couplings, $\xi_{31} \sim \xi_{13} \sim 0.3$.

3. If $m_{H^+} \gsim 600$ GeV, then both MxFV$_1$ and MxFV$_2$ models with
the corresponding $\xi_{ij} \sim 1$ are not excluded even if $m_{H^0} \sim 100$ GeV.

4. If $H^0$ decouples or
$\Phi_{FV}$ is a charged singlet, then
the constraints from $H^0$-exchanges in
$D^0 -\bar D^0$ mixing do not apply. In this case, MxFV models which have
$\xi_{32} \sim \xi_{13} \sim {\cal O}(1)$ and $m_{H^+} \sim m_W$ are also viable.

This rather surprising window of allowed ${\cal O}(1)$ MxFV
couplings can have very interesting phenomenological implications associated with
$\Phi_{FV}$ production and decays at hadron colliders.
Many aspects of this MxFV phenomenology can be understood
by noting that in MxFV models there is an
approximately conserved number $(-1)^{n_{\Phi_{FV}}+n_{t}+n_{b}}$ (similar
to the b-parity of \cite{bparity}), i.e.,
in any process involving $\Phi_{FV}$
the total number of the new FV Higgs particles
and third generation quarks is conserved modulo 2 (this symmetry is broken by terms
of order $\xi_{33} \sim V_{td} \ll 1$).
It follows that $\Phi_{FV}$ particles are produced either in pairs, mainly through
$q\bar{q} \rightarrow \Phi_{FV}\Phi_{FV}$ ($q$ is a light quark),
or in association with a single top or single b-quark via e.g.,
$g g\rightarrow \Phi_{FV} t q$, $g q \rightarrow \Phi_{FV} t,~\Phi_{FV} b$,
unless there is a b-quark in the initial state (see below).

Consider for example the MxFV$_1$ setup and assume $m_{H^0},m_{H^+} > m_t$.
In this case, the MxFV set of processes fall into 4 very well defined
categories and are therefore very distinctive at hadron colliders:
\medskip

\noindent {\bf \underline{1. $t \Phi_{FV}$ production}}
\begin{eqnarray}
&&1a)~~ dg \to t H^- \to t \bar t j, t b j + h.c. \label{Hmt}\\
&&1b)~~ ug \to t H^0 \to t t j, t \bar t j + h.c. \label{ttu1}
\end{eqnarray}
where $j$(=u or d)is a light-quark jet.
\medskip

\noindent {\bf \underline{2. $\Phi_{FV} \Phi_{FV}$ production}}
\begin{eqnarray}
&&2a)~~ u \bar u, d \bar d \to H^+ H^- \to t \bar t j j, tb jj, \bar t \bar b jj, b \bar b jj \\
&&2b)~~ u \bar u \to H^0 H^0 \to t t j j, \bar t \bar t j j, t \bar t j j \label{ttuu1}
\end{eqnarray}

\noindent {\bf \underline{3. s-channel $\Phi_{FV}$ resonance}}\\
A very interesting feature of this class of MxFV models is the
specific dynamics of resonance production of $\Phi_{FV}$.
In particular, within the class of MxFV models described in this paper,
the neutral scalar component $H^0$ {\it cannot}
be produced on resonance at hadron colliders. On the other hand,
$H^+$ can resonate via $u \bar b \to H^+$.
This can lead to a resonance peak in the invariant
mass of a $t j$ pair or, more generally, to the following 1b or 2b-jets
$H^+$-production channels:
\begin{eqnarray}
1b-tag:&& u \bar b \to H^+ \to t \bar d, \bar b j + h.c. \label{1bjet}\\
2b-tag:&& u g \to H^+ b \to t b j, b \bar b j + h.c. \label{2bjet}
\end{eqnarray}

\noindent {\bf \underline{4. t-channel $\Phi_{FV}$ exchanges}}
\begin{eqnarray}
&&3a)~~ u u \to t t +h.c. ~~(H^0 ~~{\rm exchange}) \label{tt1}\\
&&3b)~~ u d \to t b +h.c. ~~(H^+ ~~{\rm exchange})\\
&&3c)~~ u \bar u, d \bar d \to t \bar t ~~(H^+ ~\& ~H^0 ~~{\rm exchanges})\\
&&3d)~~ u \bar u \to b \bar b ~~(H^+ ~~{\rm exchanges})\\
&&3e)~~ d \bar b (dg) \to t \bar u (t \bar u b) + h.c.~~(H^+ ~~{\rm exchange})\\
&&3f)~~ ub (ug) \to u b (u b \bar b) + h.c.~~(H^+ ~~{\rm exchange})\\
&&3g)~~ ug \to t t \bar u, t \bar t u + h.c.~~(H^0 ~~{\rm exchange}) \label{ttu2}
\end{eqnarray}

An interesting limit to study is $m_t < m_{H^0} \ll m_{H^+}$ such that
$H^+$ decouples at energies relevant for the Tevatron and the LHC.
In this case we expect a noticeable signal of same-sign top-quark pairs
$pp, p \bar p \to tt + nj +X$ , where $n$ is the number of light-quark jets ($j$), from
the hard processes in (\ref{ttu1}),(\ref{ttuu1}),(\ref{tt1}) and (\ref{ttu2}).
When both top-quarks decay leptonically,
the process $pp,p \bar p \to tt + nj +X$ has a striking low background
signature of two same-sign leptons, missing energy and two b jets,
see \cite{newMxFV}. Note that, as opposed to the usual expectations,
this new $H^0$ will not be seen on resonance.

The opposite case of $m_t < m_{H^+} \ll m_{H^0}$ with a decoupled $H^0$
can also have very interesting phenomenological implications (i.e., allowing
$\xi_{32} \sim \xi_{13} \sim {\cal O}(1)$, see above).
Attractive signatures of this limiting case
will be enhanced production of $t H^-$ pairs at the LHC via
$d g \rightarrow t H^- $ [reaction (\ref{Hmt})],
and enhanced resonance production of $H^-$ via the 1 b-tag
and 2 b-tag processes in (\ref{1bjet}) and (\ref{2bjet}).
For example, we find $\sigma(pp (dg) \rightarrow t H^{-} + X) \sim 100$ [pb]
for $\xi_{31} \sim 1$ and $m_{H^-}=200$ GeV.
This is an enhancement by a factor of about 50 compared to what is considered to be the
conventional channel $bg\rightarrow t H^{-}$, e.g. in the MSSM.
In the resonance $H^-$ production channels we also expect a huge enhancement
over e.g., the MSSM or "standard" multi Higgs models for which
the leading resonance channels are $cs\rightarrow H^{\pm}, cb\rightarrow H^{\pm}$, with
couplings ${m_c\over m_W\tan\beta}V_{cs}, {m_b \tan\beta \over m_W}V_{cb} \ll 1$.

It would be of great interest to analyse these signals in the existing Tevatron data, as well as
the upcoming LHC data.
We shall leave the detailed discussion and analysis of these signals
to an upcoming work \cite{newMxFV}.

{\it Acknowledgments:} We thank D. Atwood, A. Soni and J. Feng
for useful discussions. AR is supported in part by
NSF Grants No.~PHY--0354993 and PHY--0653656.

\end{document}